\documentclass[%
 reprint,
superscriptaddress,
 amsmath,amssymb,
 aps,
]{revtex4-1}

\usepackage{graphicx}
\usepackage{dcolumn}
\usepackage{bm}

\usepackage{array}
\newcolumntype{C}[1]{>{\centering\let\newline\\\arraybackslash\hspace{0pt}}m{#1}}

\usepackage{color}
\definecolor{refcol}{RGB}{0,0,205}
\usepackage{microtype}
\usepackage[colorlinks,linkcolor=refcol,citecolor=refcol,urlcolor=refcol]{hyperref}

\newcommand {\tc} {T_c^0}
\newcommand {\ms} {m_s}

\begin{document}

\title{Universal location of the Yang-Lee edge singularity in O(N) theories} 

    \author{Andrew Connelly}
    \email{ajconne2@ncsu.edu}
    
    \author{Gregory Johnson}
    \email{Gljohns3@ncsu.edu} 
    \affiliation{Department of Physics, North Carolina State University, Raleigh, NC 27695, USA}
    
    \author{Fabian Rennecke}
    \email{frennecke@bnl.gov} 
    \affiliation{Department of Physics, Brookhaven National Laboratory, Upton, NY 11973, USA}

    \author{Vladimir V. Skokov}
    \email{VSkokov@ncsu.edu} 
    \affiliation{Department of Physics, North Carolina State University, Raleigh, NC 27695, USA}
    \affiliation{Riken-BNL Research Center, Brookhaven National Laboratory, Upton, NY 11973, USA}

\date{\today}

\begin{abstract}
We determine a previously unknown universal quantity, the location of the Yang-Lee edge singularity for the O($N$) theories in a wide range of $N$ and various dimensions. At large $N$, we reproduce the $N\to\infty$ analytical result on the location of the singularity and, additionally, we obtain the mean-field result for the location in $d=4$ dimensions. In order to capture the nonperturbative physics for arbitrary $N$, $d$ and complex-valued external fields, we use the functional renormalization group approach. 
\end{abstract}

\maketitle


\section{Introduction}

The divergence of the correlation length in the vicinity of a second order phase transition implies that the physical properties of the system are completely determined by long-range excitations. As a result, microscopic details become irrelevant and vastly different systems exhibit the same critical behavior. This enables us to group them into universality classes, which are typically defined by the symmetry and the dimensionality. This reduction instructs that by studying the simplest system within a universality class one can infer a wealth of physics describing all other members of the same class, regardless of their microscopic complexity.

Perhaps the most ubiquitous universality classes are critical O$(N)$ theories, as they describe crucial aspects of a variety of phenomena, ranging from the liquid-vapor transition of water and Curie points of magnets, to the phase diagram of QCD \cite{Pelissetto:2000ek}. A given universality class is characterized by only a few universal quantities, and
decades of theoretical and experimental research made almost all universal quantities of the relevant O$(N)$ theories known to a high precision. In the following, we argue that there is one notable exception: the location of the Yang-Lee edge singularity, a universal quantity relevant for the thermodynamic properties of the system, has not been determined so far outside of mean-field theory and the large $N$ limit. This will be remedied here. 

Near the critical point, due to the absence of a characteristic
length scale, the free energy is a homogeneous function of the relevant thermodynamic parameters, the reduced temperature, $t\propto (T-T_c)$, and the symmetry-breaking external field, $h$. The critical properties 
of the system are then quantified by a universal function of one variable: the magnetic equation of state $M=h^{1/\delta} f_G(z)$, where $z$ is the scaling variable $z=t h^{- 1/\Delta}$ with 
$\Delta=\beta\delta$.  $\beta$ and $\delta$ are universal critical exponents and $f_G(z)$ is the universal scaling function characterizing all systems within the same universality class. 

Yang and Lee have shown that the phase structure of the system is determined by the analytical structure of the equation of state for complex values of the thermodynamic parameters \cite{PhysRev.87.404,PhysRev.87.410}. Specifically, the Lee-Yang theorem states that the equation of state of O$(N)$-symmetric $\phi^4$ theories has a branch cut at purely imaginary values of $h$. A second-order (first-order) phase transition occurs when the cut pinches (crosses) the real $h$-axis. In the symmetric phase above the critical temperature, $T>T_c$, the cuts terminate at complex-conjugate branch points $\pm i h_c(T)$, known as the Yang-Lee edge singularities. Remarkably, the edge singularities are critical points themselves, characterized by a $\phi^3$ theory with purely imaginary coupling, independent of the number of field-components $N$ \cite{Fisher:1978pf}.
Near 
$h_c$, the magnetization behaves as $M\sim M_{\rm c} + (h^2-h_c^2)^{\sigma_{\rm YL}}$, with the edge critical exponent $\sigma_{\rm YL}=0.085(1)$ \cite{Gliozzi:2014jsa, An:2016lni, Zambelli:2016cbw}.

The scaling function naturally incorporates the edge singularities. The Lee-Yang theorem  
fixes the \textit{argument} of the scaling variable at the edge 
singularity $z_c=|z_c|\exp(\frac{i \pi} {2 \Delta})$. 
But while the critical exponents have been studied extensively, the location of the edge singularity, i.e.\ the {\it absolute value} $|z_c|$, has never been found for $1<d<4$ outside of the mean-field approximation or the large $N$ limit!
The importance of finding the location is difficult to overstate, as being the closest to the real axis, this singularity determines the behaviour of the equation of state for real values of the thermodynamic parameters. The high-order coefficients of a Taylor expansion  of the 
free energy around any finite value of $z$ are defined by $z_c$ and $\sigma_{\rm YL}$. This is of particular interest, e.g., for the determination of the QCD equation of state at finite density on the lattice, where a sign problem hinders direct simulations. A common strategy is to use extrapolations based on the expansion of the free energy about vanishing chemical potential \cite{Bazavov:2017dus, Borsanyi:2018grb}.
The edge singularity determines the radius of convergence of this expansion \cite{Stephanov:2006dn, Mukherjee:2019eou, Connelly:2020pno}. 

Why then was the location of the edge singularity not found before? Two of the most commonly applied methods for studying critical phenomena, the $\epsilon$-expansion and lattice simulations, are incapable of probing the Yang-Lee edge singularity. On one hand, the $\epsilon$-expansion applied to $\phi^4$ theories near four dimensions is not suitable for determining the location of the edge singularity since the critical $\phi^3$ theory in its vicinity has an upper critical dimension of six. 
This manifests itself in the appearance of non-perurbative terms~\cite{An:2017brc} when one attempts to extract the location of the edge singularity~\footnote{The determination of the location can be used to fix this non-perturbative terms and improve the widely applied parameterization of the magnetic equation of state near Ising model critical point~\cite{Nonaka:2004pg,Stephanov:2011pb,Parotto:2018pwx}.}. On the other hand, powerful, non-perturbative lattice simulations are hindered by a sign problem, as in order to extract the location of the singularity one has to perform them at imaginary values of $h$ or complex values of the temperature.  

In this letter, for the first time, we determine the previously unknown location of the Yang-Lee edge singularity for O$(N)$ symmetric $\phi^4$ theories for a wide range of $N$ in various dimensions. Additionally, we obtain the known analytic result in the large $N$ limit and, by varying the number of dimensions, reproduce the mean-field approximation. This is facilitated by using a systematic approximation of the nonperturbative functional renormalization group which enables us to determine the location $z_c$ from first principles.

\section{FRG approach to scaling for complex external fields}

To capture the inherently nonperturbative physics in the vicinity of a second order phase transition, we use the functional renormalization group (FRG) \cite{Wetterich:1992yh, Berges:2000ew, Delamotte:2003dw, Braun:2011pp, Dupuis:2020fhh}. It describes the RG flow of the scale-dependent effective action, $\Gamma_k$, as a function of a momentum scale parameter, $k$,
\begin{equation}\label{flow}
    \partial_k\Gamma_k[\phi]=\frac{1}{2} \mbox{Tr} \Bigg\{ \partial_kR_{k}\bigg(\frac{\delta^2\Gamma_k[\phi]}{\delta \phi_i \delta \phi_j}+R_{k}\bigg)^{-1} \Bigg\}, 
\end{equation}
where $R_{k}$ is the infrared regulator function which determines how the low momentum modes are screened at scale $k$.
The flow equation (\ref{flow}) prescribes the behaviour of $\Gamma_k$ between the classical action at an initial scale $k=\Lambda$ in the ultraviolet, $\Gamma_{k=\Lambda}=S$, and the desired full quantum action at $k=0$, $\Gamma_{k=0}=\Gamma$. The FRG provides a versatile realization of the Wilsonian RG and is as such well-suited to study critical physics. Both the scaling function and the critical exponents have been computed in great detail for O$(N)$ theories for real external fields with the FRG, see, e.g., \cite{Berges:1995mw, Bohr:2000gp, Litim:2001dt, Bervillier:2007rc, Braun:2007td, Braun:2008sg, Benitez:2009xg, Stokic:2009uv, Litim:2010tt, Benitez:2011xx, Rancon_2013, Defenu:2014bea, Codello:2014yfa, Eichhorn:2016hdi, Litim:2016hlb, Juttner:2017cpr, Roscher:2018ucp, Yabunaka:2018mju, DePolsi:2020pjk}.

As we alluded to above, a defining characteristic of universality is that the system is completely determined by the long-range physics of the slow critical modes. Hence, the effective action is dominated by small-momentum modes in the vicinity of a second-order phase transition. In fact, $\Gamma_k$ can be expanded systematically in powers of momentum $p^2/k^2$, which is known as the derivative expansion \cite{Morris:1994ie, Litim:2001dt}; and this expansion has been shown to be rapidly convergent within the non-perturbative regularization scheme of the FRG \cite{Balog:2019rrg}. Hence, the FRG with the derivative expansion is an exact, \textit{first-principle} method to address the Yang-Lee edge singularity of O$(N)$ critical theories.
In this letter, we use the first order derivative expansion of the scale-dependent effective action in Euclidean space,
\begin{equation}\label{LPA'}
    \Gamma_k=\int d^{d+1}x\, \bigg\{\frac{1}{2} Z_{k}(\rho)(\partial_\mu\phi)^2 + U_k(\rho)-h \sigma \bigg\}\,,
\end{equation}
where $d$ is the number of spatial dimensions, the field $\rho$ is defined as  $\rho=\phi^2/2=(\sigma^2+\pi^2)/2$, $U_k(\rho)$ is the scale dependent potential, and $h$ is the external field  explicitly breaking the O$(N)$ symmetry. $\sigma$ denotes the radial excitation, and $\pi$ the Goldstone modes. The extension to finite temperature is done by using the standard Matsubara formalism  $\int \frac{dp_0}{2\pi} \to T\sum_n$ and  $p_0 \to 2n\pi T$. Accordingly, we use the optimized regulator function $R_{k}$ for the spatial momentum modes, $R_k(\vec{p}) = Z_k (k^2-\vec{p}^{\,2}) \Theta(k^2-\vec{p}^{\, 2})$, put forward in \cite{Litim:2000ci,Litim:2001up} and refer to \cite{Litim:2001fd, Canet:2002gs, Pawlowski:2005xe, Litim:2007jb, Pawlowski:2015mlf} for detailed discussions regarding the regularization scheme. Here we comment that the commonly applied local potential approximation (LPA) which neglects the scale-dependence of the wavefunction renormalization, $Z_{\phi,k}(\rho)=1$, giving zero anomalous dimension $\eta$, is not appropriate to study the location of the edge singularity since $\eta$ is expected to be of order one in its vicinity. This necessitates the inclusion of scale-dependent wavefunction renormalization, which we assume to be field-independent, $Z_{k}(\rho)=Z_{k}$. This approximation is referred to as LPA$^\prime$. The anomalous dimension is related to the wavefunction renormalization via  $\partial_t Z_k=-\eta_k Z_k$, where we introduced 
the RG time $t=\ln(k/\Lambda)$.

To describe critical phenomena, it is helpful to look at the renormalization flow of dimensionless observables rather than the dimensionful counterparts as their magnitudes can vary wildly. The flow for the dimensionless effective potential, $\bar{U}_k(\bar{\rho})=k^{-d-1}U_k(\bar{\rho})$ where $\bar{\rho}=Z_kk^{1-d}\rho$ is the dimensionless renormalized field, can be obtained by evaluating (\ref{flow}) on uniform field configurations at fixed $\bar{\rho}$ which yields \cite{Tetradis:1993bx, Litim:2006ag}
\begin{align}\label{the Flow}
    &\partial_t \bar{U}_k(\bar{\rho})=-(d+1) \bar{U}_k+(d-1+\eta_k)\bar{\rho}\bar{U}'_k \notag \\ &+
    \frac{v_d}{2(2\pi)^d}
    \left(1-\frac{1}{d+2}\eta_{k}\right) \Bigg[\frac{N-1}{\sqrt{1+\bar{m}_\pi^2}}\bigg(1+2n_\pi\bigg) \notag \\
   & +\frac{1}{\sqrt{1+\bar{m}_\sigma^2}}\bigg(1+2n_\sigma\bigg) \Bigg]\, ,
\end{align}
where 
$v_d = S_{d-1}/d$ with $S_{d-1} = 2 \pi^{d/2} / \Gamma(d/2)$, $n_{\sigma,\pi} = n_{\sigma,\pi}(T,k)$ are the standard Bose-Einstein distribution functions, and the primes denote derivatives with respect to $\bar{\rho}$. The dimensionless renormalized masses, $\bar{m}_{\sigma,\pi}$, are given by $\displaystyle\bar{m}^2_{\pi}=\bar{U}'$ and $\displaystyle\bar{m}^2_{\sigma}=\bar{U}'+2\bar{\rho}\bar{U}''$.

In order to numerically solve (\ref{the Flow}) we consider a sixth order Taylor expansion of the dimensionless effective potential about the  scale dependent minimum $\bar{\rho}_{0,k}$: $\bar{U}_k=\sum_{i=0}^6\frac{a_{i,k}}{i!}(\bar{\rho}-\bar{\rho}_{0,k})^i$. We explicitly checked that the expansion is converged at this order.

To extract the anomalous dimension, one uses the projection \begin{equation}
    \eta_{\phi,k}=-\frac{1}{2Z_{k}}\lim_{p \to 0}\frac{\partial^2}{\partial p^2} \left[  \frac{\delta^2  \left( \partial_t \Gamma_k\right) }{\delta \pi_i(p)\delta \pi_i(-p)} \right] \,,
\end{equation}
where the choice of $i$ is arbitrary. The result is
\begin{align}\label{eta} 
    \eta_{\phi,k} &=\frac{4 v_d}{(2\pi)^d} \bar{\rho}_0 \bar{U}''(\bar{\rho}_0)^2 \\ &\times \frac{T}{k} \sum_{n=-\infty}^{\infty} \left(\frac{p_0^2}{k^2}+\bar{m}_\pi^2+1\right)^{-2} \left(\frac{p_0^2}{k^2}+\bar{m}_\sigma^2+1\right)^{-2}\, .
\notag
\end{align}
The above sum is evaluated to a closed form in practice, though its form its not insightful \cite{Paw}. It is also important to note the occurrences of $N$ and the dimension $d$ in equations (\ref{the Flow}) and (\ref{eta}). Clearly, one can easily change $N$ with no other considerations as long as $N\geq1$. As for the dimension $d$, we can safely analytically continue $d$ to any non-integer value.

To extract the scaling function $f_G(z)$ for a given $N$ and $d$, we first compute the universal critical exponents $\beta$ and $\delta$ along with the non-universal critical amplitudes $T_0$ and $H_0$. The amplitudes are defined via the replacements $ \displaystyle t \to t/T_0$ and $ \displaystyle h \to h/H_0$ so as to reproduce the scaling forms $\sigma = (-t)^\beta$ and $\sigma =  h^{1/\delta}$ at $h=0$ and $t=0$ respectively. With these conditions the magnetic equation of state 
satisfies the following normalization conditions $f_G(z=0)=1$ and $f_G(z\to -\infty) = (-z)^\beta$. 
In terms of the FRG variables~\footnote{We checked that our numerical calculations reproduce the O(4) magnetic equation of state of Ref.~\cite{Braun:2007td, Engels:2011km} for real values of the argument.} 
\begin{equation}
    f_G(z=th^{-1/\Delta})= h^{-1 /\delta} \lim_{k \to 0} \sqrt{2 \bar{\rho}_{0,k} k^{d-1} / Z_k}\,.
\end{equation}
In contrast to lattice Monte-Carlo calculations, FRG does not have a sign problem; this enables us to evaluate $f_G(z)$ for any complex $z$. In practice, we fix a real value of $t$ and numerically find $f_G(z)$ at a complex value $h=h(z)$. 
\begin{figure}[!t]
    \centering
    \includegraphics[width=0.45\textwidth]{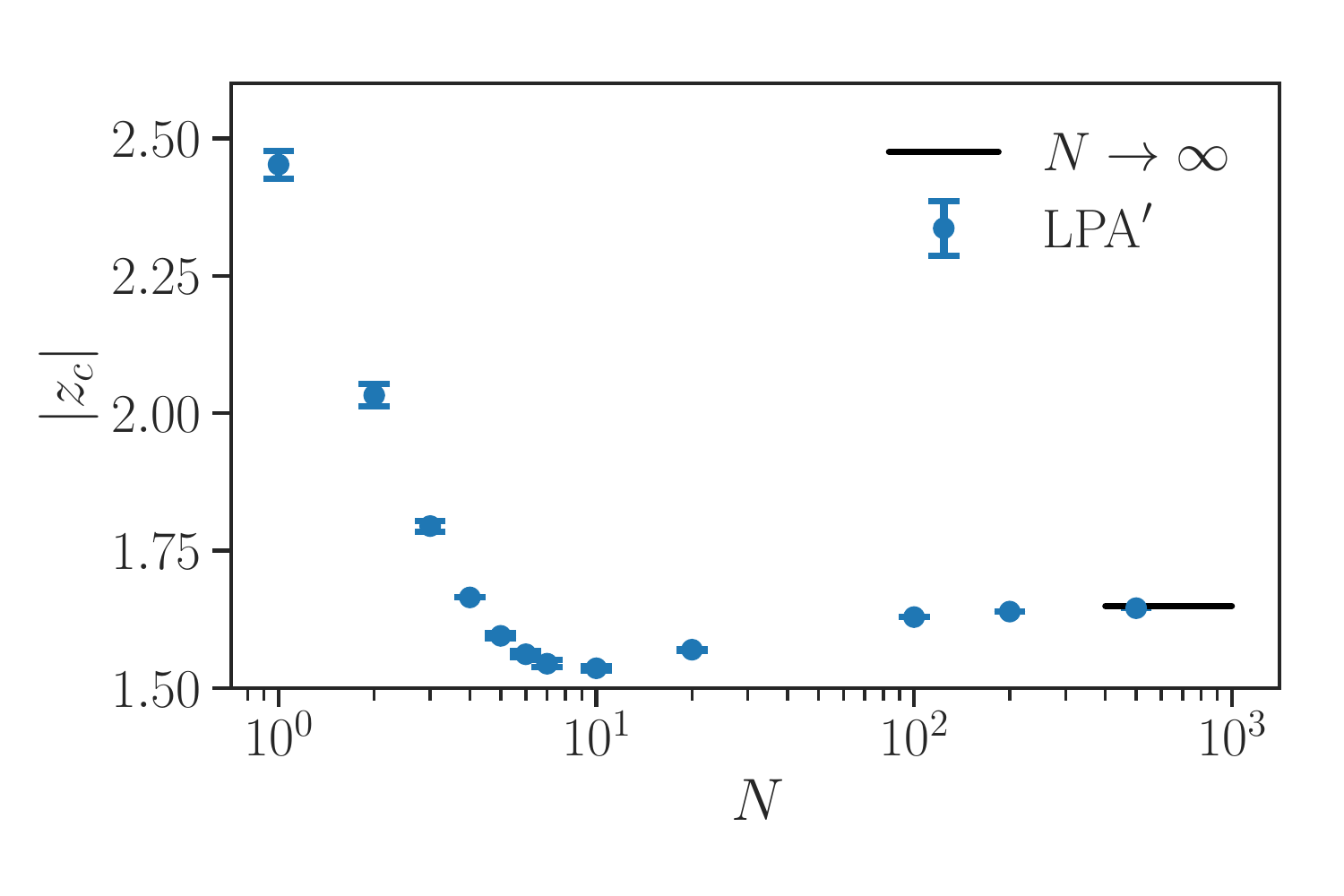}
    \caption{The magnitude of the location of the Yang-Lee edge singularity $|z_c|$  as a function of $N$ in three dimensions. The large $N$ result is indicated by the solid line.  There is a non-monotonous behaviour in the approach to large $N$ limit. }  
    \label{fig:zc}
\end{figure}

Additional care needs to be taken when approaching the edge singularity  using FRG with the Taylor expansion method. This is due to the locality of the expansion and its failure to capture first-order phase transitions. Furthermore, the edge occurs at purely imaginary values of the external field, another case for which the FRG equations become numerically challenging.  To circumvent this issue, we note that the edge can be treated as an ordinary second-order phase transition, and we can determine its location by studying the peak of the magnetic susceptibility, $\chi_\sigma\propto1/m^2_\sigma$, at a complex external field with just a small real part. Given that near the edge the order parameter $\sigma$ behaves as $\sigma-\sigma_c \sim (h^2-h_c^2)^{\sigma_{\rm YL}}$ \cite{Fisher:1978pf},
the susceptibility for the sigma field behaves as $\chi_\sigma \sim 2h \, \sigma_{\rm YL}\, (h^2-h^2_c)^{\sigma_{\rm YL}-1}$. From this form, we see that for any sufficiently small Re$(h)$, $|\chi_\sigma|$ will have a peak at some Im$(h)=h_{\text{peak}}$. Moreover, $h_\text{peak}$ is a quadratic function of Re$(h)$ for sufficiently small values of Re$(h)$. Thus, we can scan the complex $h$-plane at fixed reduced temperature $t$ along lines in the first quadrant parallel to the imaginary axis. Then $h_c$ can be determined by incrementally decreasing Re$(h)$, determining $h_{\text{peak}}$ for each increment, performing a quadratic fit to the data, and extrapolating to purely imaginary external fields. From this, we find the location of the edge as $z_c=t h_c^{-1/\Delta}$.

\begin{figure}[!t]
    \centering
    \includegraphics[width=0.45\textwidth]{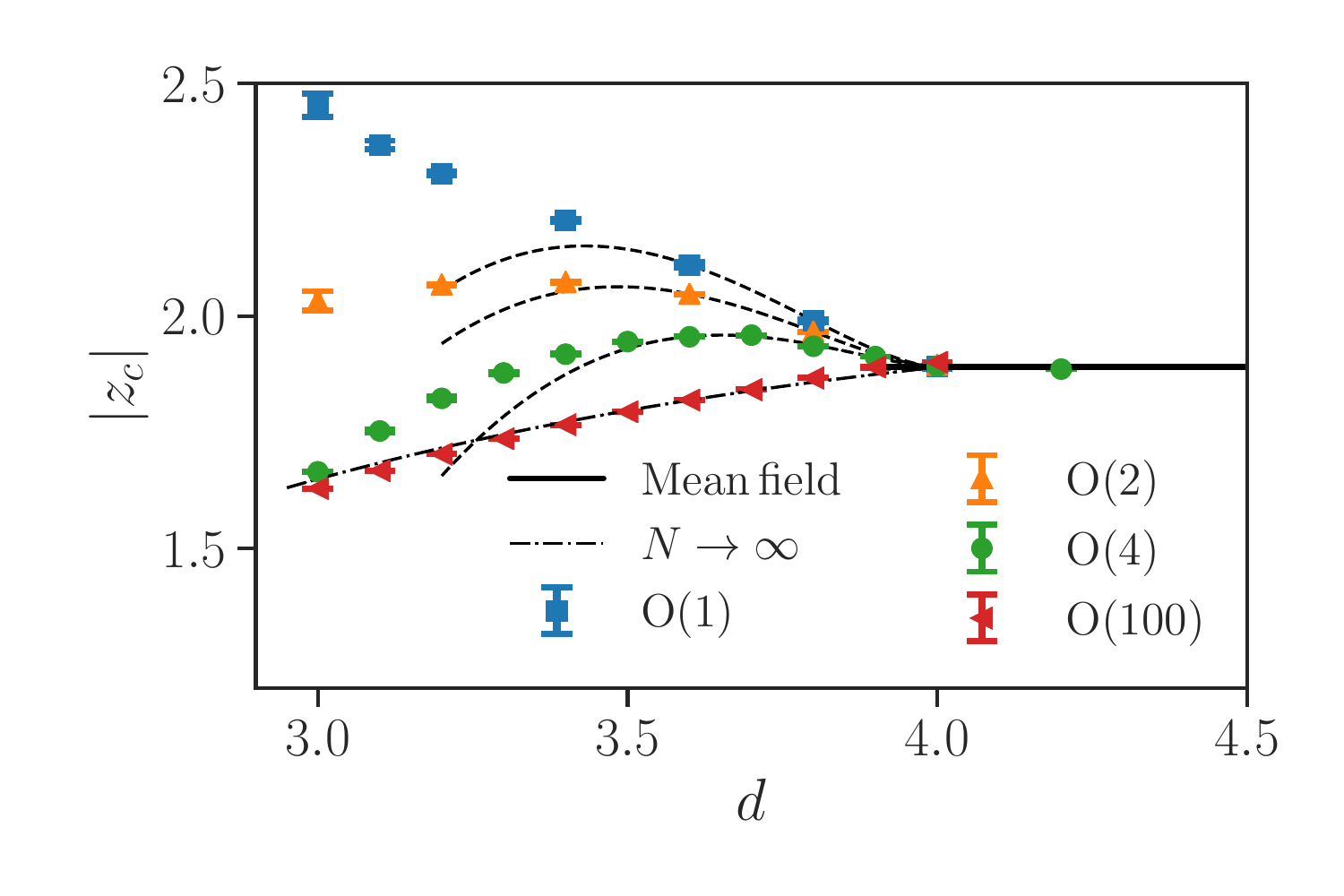}
    \caption{The magnitude of the location of the Yang-Lee edge singularity $|z_c|$  for $N=1, 2,4, 100$  as a function of the number of dimensions $d$. The solid line demonstrates the mean-field value. The dashed lines display a fit motivated by the parametrization of the equation of state for O(1)
    $|z_c| = |z^{\rm MF}_c| \left\{ 1 +  (4-d)^2 [ p_1 \ln (4-d) + p_2  ] \right\}$, see Ref.~\cite{An:2017brc}. Note that the parameter $p_2$ is non-perturbative and cannot be found using the $\epsilon$-expansion. 
    The dash-dotted line shows the analytical $N\to\infty$ result of Eq.~\eqref{Eq:LargeN} with $\gamma=2/(d-2)$.    
    } 
    \label{fig:zcd}
\end{figure}

There are several sources of error in this work, presumably the largest being the systematic error which results from working in LPA$^\prime$. 
One manifestation of the systematic error can be the violation of the hyperscaling relation between the exponents: $2-\eta=d(\delta-1)/(\delta+1)$. For our work, we find this violation to be at the sub-permille level. 
Regarding the critical exponents themselves, e.g.\ for $d=3$ and $N=1(2)$, the disagreement with well-known results from Ref.~\cite{Kos:2016ysd} for $\Delta$ is about 0.9\% (1.9\%). 
We note that a potentially large anomalous dimension close to the edge singularity indicates nontrivial momentum dependencies in the system. These can be captured by higher orders in the derivative expansion. A systematic improvement of our truncation in this direction, which will significantly reduce the systematic error, is deferred to future work~\cite{CRS2020}.

Other error sources are the implementation of a finite infrared scale $k_0$ at which the RG flows are terminated and the fits to obtain the exponents and the value of $h_c$. 
One can choose a $k_0$ such that the continued running of the flows below this value contributes negligibly to the error. 
For this work, we chose a common $k_0$ for all $N$ and $d$ such that the error is negligibly small. The error from the fits for the exponents and extrapolation to purely imaginary values of $h$ are much smaller than the truncation error and we do not report them here. The error bars seen in Figures \ref{fig:zc} and \ref{fig:zcd} are then given by the truncation error, which we measure as the difference between the result at $5^{th}$ and $6^{th}$ order Taylor expansions of the effective potential.

\section{Results and discussions} 

Our results for the location of the Yang-Lee edge singularity, $z_c$, are shown in Figure \ref{fig:zc} for $d=3$ and numerous $N$, and in Figure \ref{fig:zcd} for $N=1, 2,4,100$ and various $d$. Analytically, $z_c$ is only known in two limiting cases: $N\rightarrow \infty$ and the mean-field approximation. For both, the magnetic equation of state can be written in the following form (see e.g. ~\cite{Brezin:1972se,Moshe:2003xn,Almasi:2016gcn}) 
\begin{equation}
    f_G(z) \left[z + f_G^2(z)\right]^\gamma = 1,  
\end{equation}
where $\gamma=1$ ($\gamma=\frac{2}{d-2}$) for mean-field (large $N$ in $d<4$ dimensions). Due to its nature as a branch point, the location of the Yang-Lee edge singularity can be found from the condition that the derivative of the inverse function is zero, $z'(f_G)=0$. This leads to 
\begin{equation}
    |z_c| = (2 \gamma + 1)  \left(2 \gamma\right)^{-\frac{2 \gamma}{2 \gamma+1}}  , 
    \quad {\rm Arg}\,  z_c  = \pm \pi (2\gamma+1)^{-1}  \,. 
    \label{Eq:LargeN}
\end{equation}
Thus in mean-field approximation (in the large $N$ limit, $d=3$)  one gets $|z^{\rm MF}_c| \approx 1.8899$ ($|z^{\infty}_c| \approx 1.6494$). 

The fact that our numerical calculations reproduce the analytical results for $|z_c|$
is highly non-trivial in that it provides a direct test of our FRG approach and the method of extracting the location of the singularity. Figure \ref{fig:zc} demonstrates the converge of $|z_c|$ to the large $N$ value. 
We note that $|z_c|$ displays some non-monotonous behavior as it dips below its large $N$ value before approaching the limit from below. For the convergence of $|z_c|$ to its mean field result, see Figure \ref{fig:zcd}. We also see here non-monotonous behavior as the dimension is varied. Our numerical calculation for $N=100$ faithfully reproduces the infinite $N$ result in arbitrary number of dimensions $3<d<4$. 
In three dimensions, 
the values of $|z_c|$ are 
$2.452 \pm 0.025$, $2.032 \pm 0.021$ and $1.665 \pm 10^{-3}$
for $N=$1, 2, 4  correspondingly.  

To give an explicit example for importance of the edge singularity, we determine the asymptotic form of the coefficients of the Taylor expansion of $f_G(z)$. This can be done solely based on its analytic structure in the complex plane. For the sake of simplicity, we  
consider the expansion near the origin. In this case, the theorem of Darboux states (see e.g.~Ref.~\cite{henrici1991applied}) that the $n$-th expansion coefficient for sufficiently large $n$ is 
\begin{align}
f_G^{(n)} \sim 2 B_0 |z_c|^{-n} \frac{n^{\sigma_{\rm YL}-1}}{\Gamma(\sigma_{\rm YL})}  \cos \left(\beta_0 -  \frac{\pi n}{2\Delta} \right),  
\end{align}
where $B_0$ and $\beta_0$ are defined as the value of the analytic part of the magnetization at the singularity $B_0 \exp (i \beta_0) = \lim\limits_{z\to z_c}   (1-z/z_c)^{-\sigma_{\rm YL}} ( f_G(z)-f_G(z_c))$. 
This explicitly demonstrates that the expansion coefficients are uniquely defined by $|z_c|$, $\Delta$ and $\sigma_{\rm YL}$. 

Another immediate application of our results is to lattice QCD calculations~\cite{Mukherjee:2019eou}. As we alluded to before, due to the sign problem, lattice calculations can not be effectively performed at non-zero baryon chemical potential/density $\mu\ne 0$.
Nevertheless, extrapolations to finite $\mu$ based on a Taylor series expansion of, e.g., the pressure about $\mu=0$ are used. In the crossover region, this series expansion has a finite radius of convergence which is defined by $z_c$ and a non-universal map of $T$, $\mu$, and $m$ to the scaling variable $z$,
\begin{align}
  z(T,\mu) &= z_0 \left( \frac{m_{u,d}}{\ms}\right)^{-\frac{1}{\Delta}} 
  \left[ \frac{T-\tc}{\tc} + \kappa^B_2 \left( \frac{\mu}{\tc} \right)^2  + \dots \right] \,.
  \notag
\end{align}
The non-universal parameters ($\tc$, $\kappa_2^B$, and $z_0$) for this map were computed in lattice QCD
see Refs.~\cite{Bazavov:2017dus,Bazavov:2018mes,Borsanyi:2019hsj,Mukherjee:2019eou}. 
This, complemented by our result, shows that the radius of convergence $R=|\mu_c|$ defined by $z(T,\mu_c)=z_c$ is 
$\ge 420$ MeV for the O(4) universality class and $\ge 460$ MeV for the O(2) universality class (which is of relevance for staggered fermions). These numbers are consistent with the ratio test applied to the Taylor series expansion in Refs.~\cite{Bazavov:2017dus}. More importantly, they are well below the latest results for the location of a critical endpoint in the QCD phase diagram~\cite{Fischer:2018sdj, Fu:2019hdw, Gao:2020qsj}. This highlights that the edge singularity, not the critical endpoint, is the crucial quantity for explorations of the QCD phase structure  by lattice QCD.

\section{Conclusions}

In this letter we have provided the first results on a previously unknown universal quantity for O$(N)$-like theories. We determined the location of the Yang-Lee edge singularity for the O($N$) universality class for a wide range of $N$ and dimensions $d$. Our numerical calculations reproduce known analytical results for $z_c$ in the infinite $N$ limit and in the mean-field approximation ($d=4$). The computation described here can readily be adapted for the precise study of the edge singularity in other universality classes as well.

\begin{acknowledgments}
We thank 
B.\ Friman,
A.\ Kemper, 
J.\ M.\ Pawlowski,
R.\ Pisarski, 
K.\ Redlich, 
T.\ Schaefer,
M.\ Stephanov, 
M.\ \"Unsal, and 
N.\ Wink
for illuminating discussions. 
V.S.\ is thankful to S.\ Mukherjee for encouragement and discussions which eventually led to this work.   
V.S.\ acknowledges support by the DOE Office of Nuclear Physics through Grant No.\ DE-SC0020081. 
F.R.\ is supported by the U.S. Department of Energy under contract DE-SC0012704.
V.S.\ thanks
the ExtreMe Matter Institute EMMI (GSI Helmholtzzentrum f\"ur Schwerionenforschung, Darmstadt, Germany) for
partial support and hospitality. 
\end{acknowledgments}

\bibliography{edgy}

\providecommand{\noopsort}[1]{}\providecommand{\singleletter}[1]{#1}%
\begin{thebibliography}{66}%
\makeatletter
\providecommand \@ifxundefined [1]{%
 \@ifx{#1\undefined}
}%
\providecommand \@ifnum [1]{%
 \ifnum #1\expandafter \@firstoftwo
 \else \expandafter \@secondoftwo
 \fi
}%
\providecommand \@ifx [1]{%
 \ifx #1\expandafter \@firstoftwo
 \else \expandafter \@secondoftwo
 \fi
}%
\providecommand \natexlab [1]{#1}%
\providecommand \enquote  [1]{``#1''}%
\providecommand \bibnamefont  [1]{#1}%
\providecommand \bibfnamefont [1]{#1}%
\providecommand \citenamefont [1]{#1}%
\providecommand \href@noop [0]{\@secondoftwo}%
\providecommand \href [0]{\begingroup \@sanitize@url \@href}%
\providecommand \@href[1]{\@@startlink{#1}\@@href}%
\providecommand \@@href[1]{\endgroup#1\@@endlink}%
\providecommand \@sanitize@url [0]{\catcode `\\12\catcode `\$12\catcode
  `\&12\catcode `\#12\catcode `\^12\catcode `\_12\catcode `\%12\relax}%
\providecommand \@@startlink[1]{}%
\providecommand \@@endlink[0]{}%
\providecommand \url  [0]{\begingroup\@sanitize@url \@url }%
\providecommand \@url [1]{\endgroup\@href {#1}{\urlprefix }}%
\providecommand \urlprefix  [0]{URL }%
\providecommand \Eprint [0]{\href }%
\providecommand \doibase [0]{http://dx.doi.org/}%
\providecommand \selectlanguage [0]{\@gobble}%
\providecommand \bibinfo  [0]{\@secondoftwo}%
\providecommand \bibfield  [0]{\@secondoftwo}%
\providecommand \translation [1]{[#1]}%
\providecommand \BibitemOpen [0]{}%
\providecommand \bibitemStop [0]{}%
\providecommand \bibitemNoStop [0]{.\EOS\space}%
\providecommand \EOS [0]{\spacefactor3000\relax}%
\providecommand \BibitemShut  [1]{\csname bibitem#1\endcsname}%
\let\auto@bib@innerbib\@empty
\bibitem [{\citenamefont {Pelissetto}\ and\ \citenamefont
  {Vicari}(2002)}]{Pelissetto:2000ek}%
  \BibitemOpen
  \bibfield  {author} {\bibinfo {author} {\bibfnamefont {A.}~\bibnamefont
  {Pelissetto}}\ and\ \bibinfo {author} {\bibfnamefont {E.}~\bibnamefont
  {Vicari}},\ }\href {\doibase 10.1016/S0370-1573(02)00219-3} {\bibfield
  {journal} {\bibinfo  {journal} {Phys. Rept.}\ }\textbf {\bibinfo {volume}
  {368}},\ \bibinfo {pages} {549} (\bibinfo {year} {2002})},\ \Eprint
  {http://arxiv.org/abs/cond-mat/0012164} {arXiv:cond-mat/0012164} \BibitemShut
  {NoStop}%
\bibitem [{\citenamefont {Yang}\ and\ \citenamefont
  {Lee}(1952)}]{PhysRev.87.404}%
  \BibitemOpen
  \bibfield  {author} {\bibinfo {author} {\bibfnamefont {C.~N.}\ \bibnamefont
  {Yang}}\ and\ \bibinfo {author} {\bibfnamefont {T.~D.}\ \bibnamefont {Lee}},\
  }\href {\doibase 10.1103/PhysRev.87.404} {\bibfield  {journal} {\bibinfo
  {journal} {Phys. Rev.}\ }\textbf {\bibinfo {volume} {87}},\ \bibinfo {pages}
  {404} (\bibinfo {year} {1952})}\BibitemShut {NoStop}%
\bibitem [{\citenamefont {Lee}\ and\ \citenamefont
  {Yang}(1952)}]{PhysRev.87.410}%
  \BibitemOpen
  \bibfield  {author} {\bibinfo {author} {\bibfnamefont {T.~D.}\ \bibnamefont
  {Lee}}\ and\ \bibinfo {author} {\bibfnamefont {C.~N.}\ \bibnamefont {Yang}},\
  }\href {\doibase 10.1103/PhysRev.87.410} {\bibfield  {journal} {\bibinfo
  {journal} {Phys. Rev.}\ }\textbf {\bibinfo {volume} {87}},\ \bibinfo {pages}
  {410} (\bibinfo {year} {1952})}\BibitemShut {NoStop}%
\bibitem [{\citenamefont {Fisher}(1978)}]{Fisher:1978pf}%
  \BibitemOpen
  \bibfield  {author} {\bibinfo {author} {\bibfnamefont {M.~E.}\ \bibnamefont
  {Fisher}},\ }\href {\doibase 10.1103/PhysRevLett.40.1610} {\bibfield
  {journal} {\bibinfo  {journal} {Phys. Rev. Lett.}\ }\textbf {\bibinfo
  {volume} {40}},\ \bibinfo {pages} {1610} (\bibinfo {year}
  {1978})}\BibitemShut {NoStop}%
\bibitem [{\citenamefont {Gliozzi}\ and\ \citenamefont
  {Rago}(2014)}]{Gliozzi:2014jsa}%
  \BibitemOpen
  \bibfield  {author} {\bibinfo {author} {\bibfnamefont {F.}~\bibnamefont
  {Gliozzi}}\ and\ \bibinfo {author} {\bibfnamefont {A.}~\bibnamefont {Rago}},\
  }\href {\doibase 10.1007/JHEP10(2014)042} {\bibfield  {journal} {\bibinfo
  {journal} {JHEP}\ }\textbf {\bibinfo {volume} {10}},\ \bibinfo {pages} {042}
  (\bibinfo {year} {2014})},\ \Eprint {http://arxiv.org/abs/1403.6003}
  {arXiv:1403.6003 [hep-th]} \BibitemShut {NoStop}%
\bibitem [{\citenamefont {An}\ \emph {et~al.}(2016)\citenamefont {An},
  \citenamefont {Mesterhazy},\ and\ \citenamefont {Stephanov}}]{An:2016lni}%
  \BibitemOpen
  \bibfield  {author} {\bibinfo {author} {\bibfnamefont {X.}~\bibnamefont
  {An}}, \bibinfo {author} {\bibfnamefont {D.}~\bibnamefont {Mesterhazy}}, \
  and\ \bibinfo {author} {\bibfnamefont {M.~A.}\ \bibnamefont {Stephanov}},\
  }\href {\doibase 10.1007/JHEP07(2016)041} {\bibfield  {journal} {\bibinfo
  {journal} {JHEP}\ }\textbf {\bibinfo {volume} {07}},\ \bibinfo {pages} {041}
  (\bibinfo {year} {2016})},\ \Eprint {http://arxiv.org/abs/1605.06039}
  {arXiv:1605.06039 [hep-th]} \BibitemShut {NoStop}%
\bibitem [{\citenamefont {Zambelli}\ and\ \citenamefont
  {Zanusso}(2017)}]{Zambelli:2016cbw}%
  \BibitemOpen
  \bibfield  {author} {\bibinfo {author} {\bibfnamefont {L.}~\bibnamefont
  {Zambelli}}\ and\ \bibinfo {author} {\bibfnamefont {O.}~\bibnamefont
  {Zanusso}},\ }\href {\doibase 10.1103/PhysRevD.95.085001} {\bibfield
  {journal} {\bibinfo  {journal} {Phys. Rev. D}\ }\textbf {\bibinfo {volume}
  {95}},\ \bibinfo {pages} {085001} (\bibinfo {year} {2017})},\ \Eprint
  {http://arxiv.org/abs/1612.08739} {arXiv:1612.08739 [hep-th]} \BibitemShut
  {NoStop}%
\bibitem [{\citenamefont {Bazavov}\ \emph {et~al.}(2017)\citenamefont {Bazavov}
  \emph {et~al.}}]{Bazavov:2017dus}%
  \BibitemOpen
  \bibfield  {author} {\bibinfo {author} {\bibfnamefont {A.}~\bibnamefont
  {Bazavov}} \emph {et~al.},\ }\href {\doibase 10.1103/PhysRevD.95.054504}
  {\bibfield  {journal} {\bibinfo  {journal} {Phys. Rev.}\ }\textbf {\bibinfo
  {volume} {D95}},\ \bibinfo {pages} {054504} (\bibinfo {year} {2017})},\
  \Eprint {http://arxiv.org/abs/1701.04325} {arXiv:1701.04325 [hep-lat]}
  \BibitemShut {NoStop}%
\bibitem [{\citenamefont {Borsanyi}\ \emph {et~al.}(2018)\citenamefont
  {Borsanyi}, \citenamefont {Fodor}, \citenamefont {Guenther}, \citenamefont
  {Katz}, \citenamefont {Szabo}, \citenamefont {Pasztor}, \citenamefont
  {Portillo},\ and\ \citenamefont {Ratti}}]{Borsanyi:2018grb}%
  \BibitemOpen
  \bibfield  {author} {\bibinfo {author} {\bibfnamefont {S.}~\bibnamefont
  {Borsanyi}}, \bibinfo {author} {\bibfnamefont {Z.}~\bibnamefont {Fodor}},
  \bibinfo {author} {\bibfnamefont {J.~N.}\ \bibnamefont {Guenther}}, \bibinfo
  {author} {\bibfnamefont {S.~K.}\ \bibnamefont {Katz}}, \bibinfo {author}
  {\bibfnamefont {K.~K.}\ \bibnamefont {Szabo}}, \bibinfo {author}
  {\bibfnamefont {A.}~\bibnamefont {Pasztor}}, \bibinfo {author} {\bibfnamefont
  {I.}~\bibnamefont {Portillo}}, \ and\ \bibinfo {author} {\bibfnamefont
  {C.}~\bibnamefont {Ratti}},\ }\href {\doibase 10.1007/JHEP10(2018)205}
  {\bibfield  {journal} {\bibinfo  {journal} {JHEP}\ }\textbf {\bibinfo
  {volume} {10}},\ \bibinfo {pages} {205} (\bibinfo {year} {2018})},\ \Eprint
  {http://arxiv.org/abs/1805.04445} {arXiv:1805.04445 [hep-lat]} \BibitemShut
  {NoStop}%
\bibitem [{\citenamefont {Stephanov}(2006)}]{Stephanov:2006dn}%
  \BibitemOpen
  \bibfield  {author} {\bibinfo {author} {\bibfnamefont {M.}~\bibnamefont
  {Stephanov}},\ }\href {\doibase 10.1103/PhysRevD.73.094508} {\bibfield
  {journal} {\bibinfo  {journal} {Phys. Rev. D}\ }\textbf {\bibinfo {volume}
  {73}},\ \bibinfo {pages} {094508} (\bibinfo {year} {2006})},\ \Eprint
  {http://arxiv.org/abs/hep-lat/0603014} {arXiv:hep-lat/0603014} \BibitemShut
  {NoStop}%
\bibitem [{\citenamefont {Mukherjee}\ and\ \citenamefont
  {Skokov}(2019)}]{Mukherjee:2019eou}%
  \BibitemOpen
  \bibfield  {author} {\bibinfo {author} {\bibfnamefont {S.}~\bibnamefont
  {Mukherjee}}\ and\ \bibinfo {author} {\bibfnamefont {V.}~\bibnamefont
  {Skokov}},\ }\href@noop {} {\  (\bibinfo {year} {2019})},\ \Eprint
  {http://arxiv.org/abs/1909.04639} {arXiv:1909.04639 [hep-ph]} \BibitemShut
  {NoStop}%
\bibitem [{\citenamefont {Connelly}\ \emph {et~al.}(2020)\citenamefont
  {Connelly}, \citenamefont {Johnson}, \citenamefont {Mukherjee},\ and\
  \citenamefont {Skokov}}]{Connelly:2020pno}%
  \BibitemOpen
  \bibfield  {author} {\bibinfo {author} {\bibfnamefont {A.}~\bibnamefont
  {Connelly}}, \bibinfo {author} {\bibfnamefont {G.}~\bibnamefont {Johnson}},
  \bibinfo {author} {\bibfnamefont {S.}~\bibnamefont {Mukherjee}}, \ and\
  \bibinfo {author} {\bibfnamefont {V.}~\bibnamefont {Skokov}},\ }in\
  \href@noop {} {\emph {\bibinfo {booktitle} {{28th International Conference on
  Ultrarelativistic Nucleus-Nucleus Collisions}}}}\ (\bibinfo {year} {2020})\
  \Eprint {http://arxiv.org/abs/2004.05095} {arXiv:2004.05095 [hep-ph]}
  \BibitemShut {NoStop}%
\bibitem [{\citenamefont {An}\ \emph {et~al.}(2018)\citenamefont {An},
  \citenamefont {Mesterhazy},\ and\ \citenamefont {Stephanov}}]{An:2017brc}%
  \BibitemOpen
  \bibfield  {author} {\bibinfo {author} {\bibfnamefont {X.}~\bibnamefont
  {An}}, \bibinfo {author} {\bibfnamefont {D.}~\bibnamefont {Mesterhazy}}, \
  and\ \bibinfo {author} {\bibfnamefont {M.~A.}\ \bibnamefont {Stephanov}},\
  }\href {\doibase 10.1088/1742-5468/aaac4a} {\bibfield  {journal} {\bibinfo
  {journal} {J. Stat. Mech.}\ }\textbf {\bibinfo {volume} {1803}},\ \bibinfo
  {pages} {033207} (\bibinfo {year} {2018})},\ \Eprint
  {http://arxiv.org/abs/1707.06447} {arXiv:1707.06447 [hep-th]} \BibitemShut
  {NoStop}%
\bibitem [{Note1()}]{Note1}%
  \BibitemOpen
  \bibinfo {note} {The determination of the location can be used to fix this
  non-perturbative terms and improve the widely applied parameterization of the
  magnetic equation of state near Ising model critical point~\cite
  {Nonaka:2004pg,Stephanov:2011pb,Parotto:2018pwx}.}\BibitemShut {Stop}%
\bibitem [{\citenamefont {Wetterich}(1993)}]{Wetterich:1992yh}%
  \BibitemOpen
  \bibfield  {author} {\bibinfo {author} {\bibfnamefont {C.}~\bibnamefont
  {Wetterich}},\ }\href@noop {} {\bibfield  {journal} {\bibinfo  {journal}
  {Phys. Lett.}\ }\textbf {\bibinfo {volume} {B301}},\ \bibinfo {pages} {90}
  (\bibinfo {year} {1993})}\BibitemShut {NoStop}%
\bibitem [{\citenamefont {Berges}\ \emph {et~al.}(2002)\citenamefont {Berges},
  \citenamefont {Tetradis},\ and\ \citenamefont {Wetterich}}]{Berges:2000ew}%
  \BibitemOpen
  \bibfield  {author} {\bibinfo {author} {\bibfnamefont {J.}~\bibnamefont
  {Berges}}, \bibinfo {author} {\bibfnamefont {N.}~\bibnamefont {Tetradis}}, \
  and\ \bibinfo {author} {\bibfnamefont {C.}~\bibnamefont {Wetterich}},\ }\href
  {\doibase 10.1016/S0370-1573(01)00098-9} {\bibfield  {journal} {\bibinfo
  {journal} {Phys. Rept.}\ }\textbf {\bibinfo {volume} {363}},\ \bibinfo
  {pages} {223} (\bibinfo {year} {2002})},\ \Eprint
  {http://arxiv.org/abs/hep-ph/0005122} {arXiv:hep-ph/0005122 [hep-ph]}
  \BibitemShut {NoStop}%
\bibitem [{\citenamefont {Delamotte}\ \emph {et~al.}(2004)\citenamefont
  {Delamotte}, \citenamefont {Mouhanna},\ and\ \citenamefont
  {Tissier}}]{Delamotte:2003dw}%
  \BibitemOpen
  \bibfield  {author} {\bibinfo {author} {\bibfnamefont {B.}~\bibnamefont
  {Delamotte}}, \bibinfo {author} {\bibfnamefont {D.}~\bibnamefont {Mouhanna}},
  \ and\ \bibinfo {author} {\bibfnamefont {M.}~\bibnamefont {Tissier}},\ }\href
  {\doibase 10.1103/PhysRevB.69.134413} {\bibfield  {journal} {\bibinfo
  {journal} {Phys. Rev. B}\ }\textbf {\bibinfo {volume} {69}},\ \bibinfo
  {pages} {134413} (\bibinfo {year} {2004})},\ \Eprint
  {http://arxiv.org/abs/cond-mat/0309101} {arXiv:cond-mat/0309101} \BibitemShut
  {NoStop}%
\bibitem [{\citenamefont {Braun}(2012)}]{Braun:2011pp}%
  \BibitemOpen
  \bibfield  {author} {\bibinfo {author} {\bibfnamefont {J.}~\bibnamefont
  {Braun}},\ }\href {\doibase 10.1088/0954-3899/39/3/033001} {\bibfield
  {journal} {\bibinfo  {journal} {J. Phys. G}\ }\textbf {\bibinfo {volume}
  {39}},\ \bibinfo {pages} {033001} (\bibinfo {year} {2012})},\ \Eprint
  {http://arxiv.org/abs/1108.4449} {arXiv:1108.4449 [hep-ph]} \BibitemShut
  {NoStop}%
\bibitem [{\citenamefont {Dupuis}\ \emph {et~al.}(2020)\citenamefont {Dupuis},
  \citenamefont {Canet}, \citenamefont {Eichhorn}, \citenamefont {Metzner},
  \citenamefont {Pawlowski}, \citenamefont {Tissier},\ and\ \citenamefont
  {Wschebor}}]{Dupuis:2020fhh}%
  \BibitemOpen
  \bibfield  {author} {\bibinfo {author} {\bibfnamefont {N.}~\bibnamefont
  {Dupuis}}, \bibinfo {author} {\bibfnamefont {L.}~\bibnamefont {Canet}},
  \bibinfo {author} {\bibfnamefont {A.}~\bibnamefont {Eichhorn}}, \bibinfo
  {author} {\bibfnamefont {W.}~\bibnamefont {Metzner}}, \bibinfo {author}
  {\bibfnamefont {J.}~\bibnamefont {Pawlowski}}, \bibinfo {author}
  {\bibfnamefont {M.}~\bibnamefont {Tissier}}, \ and\ \bibinfo {author}
  {\bibfnamefont {N.}~\bibnamefont {Wschebor}},\ }\href@noop {} {\  (\bibinfo
  {year} {2020})},\ \Eprint {http://arxiv.org/abs/2006.04853} {arXiv:2006.04853
  [cond-mat.stat-mech]} \BibitemShut {NoStop}%
\bibitem [{\citenamefont {Berges}\ \emph {et~al.}(1996)\citenamefont {Berges},
  \citenamefont {Tetradis},\ and\ \citenamefont {Wetterich}}]{Berges:1995mw}%
  \BibitemOpen
  \bibfield  {author} {\bibinfo {author} {\bibfnamefont {J.}~\bibnamefont
  {Berges}}, \bibinfo {author} {\bibfnamefont {N.}~\bibnamefont {Tetradis}}, \
  and\ \bibinfo {author} {\bibfnamefont {C.}~\bibnamefont {Wetterich}},\ }\href
  {\doibase 10.1103/PhysRevLett.77.873} {\bibfield  {journal} {\bibinfo
  {journal} {Phys. Rev. Lett.}\ }\textbf {\bibinfo {volume} {77}},\ \bibinfo
  {pages} {873} (\bibinfo {year} {1996})},\ \Eprint
  {http://arxiv.org/abs/hep-th/9507159} {arXiv:hep-th/9507159} \BibitemShut
  {NoStop}%
\bibitem [{\citenamefont {Bohr}\ \emph {et~al.}(2001)\citenamefont {Bohr},
  \citenamefont {Schaefer},\ and\ \citenamefont {Wambach}}]{Bohr:2000gp}%
  \BibitemOpen
  \bibfield  {author} {\bibinfo {author} {\bibfnamefont {O.}~\bibnamefont
  {Bohr}}, \bibinfo {author} {\bibfnamefont {B.}~\bibnamefont {Schaefer}}, \
  and\ \bibinfo {author} {\bibfnamefont {J.}~\bibnamefont {Wambach}},\ }\href
  {\doibase 10.1142/S0217751X0100502X} {\bibfield  {journal} {\bibinfo
  {journal} {Int. J. Mod. Phys. A}\ }\textbf {\bibinfo {volume} {16}},\
  \bibinfo {pages} {3823} (\bibinfo {year} {2001})},\ \Eprint
  {http://arxiv.org/abs/hep-ph/0007098} {arXiv:hep-ph/0007098} \BibitemShut
  {NoStop}%
\bibitem [{\citenamefont {Litim}(2001{\natexlab{a}})}]{Litim:2001dt}%
  \BibitemOpen
  \bibfield  {author} {\bibinfo {author} {\bibfnamefont {D.~F.}\ \bibnamefont
  {Litim}},\ }\href {\doibase 10.1088/1126-6708/2001/11/059} {\bibfield
  {journal} {\bibinfo  {journal} {JHEP}\ }\textbf {\bibinfo {volume} {11}},\
  \bibinfo {pages} {059} (\bibinfo {year} {2001}{\natexlab{a}})},\ \Eprint
  {http://arxiv.org/abs/hep-th/0111159} {arXiv:hep-th/0111159} \BibitemShut
  {NoStop}%
\bibitem [{\citenamefont {Bervillier}\ \emph {et~al.}(2007)\citenamefont
  {Bervillier}, \citenamefont {Juttner},\ and\ \citenamefont
  {Litim}}]{Bervillier:2007rc}%
  \BibitemOpen
  \bibfield  {author} {\bibinfo {author} {\bibfnamefont {C.}~\bibnamefont
  {Bervillier}}, \bibinfo {author} {\bibfnamefont {A.}~\bibnamefont {Juttner}},
  \ and\ \bibinfo {author} {\bibfnamefont {D.~F.}\ \bibnamefont {Litim}},\
  }\href {\doibase 10.1016/j.nuclphysb.2007.03.036} {\bibfield  {journal}
  {\bibinfo  {journal} {Nucl. Phys. B}\ }\textbf {\bibinfo {volume} {783}},\
  \bibinfo {pages} {213} (\bibinfo {year} {2007})},\ \Eprint
  {http://arxiv.org/abs/hep-th/0701172} {arXiv:hep-th/0701172} \BibitemShut
  {NoStop}%
\bibitem [{\citenamefont {Braun}\ and\ \citenamefont
  {Klein}(2008)}]{Braun:2007td}%
  \BibitemOpen
  \bibfield  {author} {\bibinfo {author} {\bibfnamefont {J.}~\bibnamefont
  {Braun}}\ and\ \bibinfo {author} {\bibfnamefont {B.}~\bibnamefont {Klein}},\
  }\href {\doibase 10.1103/PhysRevD.77.096008} {\bibfield  {journal} {\bibinfo
  {journal} {Phys. Rev. D}\ }\textbf {\bibinfo {volume} {77}},\ \bibinfo
  {pages} {096008} (\bibinfo {year} {2008})},\ \Eprint
  {http://arxiv.org/abs/0712.3574} {arXiv:0712.3574 [hep-th]} \BibitemShut
  {NoStop}%
\bibitem [{\citenamefont {Braun}\ and\ \citenamefont
  {Klein}(2009)}]{Braun:2008sg}%
  \BibitemOpen
  \bibfield  {author} {\bibinfo {author} {\bibfnamefont {J.}~\bibnamefont
  {Braun}}\ and\ \bibinfo {author} {\bibfnamefont {B.}~\bibnamefont {Klein}},\
  }\href {\doibase 10.1140/epjc/s10052-009-1098-8} {\bibfield  {journal}
  {\bibinfo  {journal} {Eur. Phys. J. C}\ }\textbf {\bibinfo {volume} {63}},\
  \bibinfo {pages} {443} (\bibinfo {year} {2009})},\ \Eprint
  {http://arxiv.org/abs/0810.0857} {arXiv:0810.0857 [hep-ph]} \BibitemShut
  {NoStop}%
\bibitem [{\citenamefont {Benitez}\ \emph {et~al.}(2009)\citenamefont
  {Benitez}, \citenamefont {Blaizot}, \citenamefont {Chate}, \citenamefont
  {Delamotte}, \citenamefont {Mendez-Galain},\ and\ \citenamefont
  {Wschebor}}]{Benitez:2009xg}%
  \BibitemOpen
  \bibfield  {author} {\bibinfo {author} {\bibfnamefont {F.}~\bibnamefont
  {Benitez}}, \bibinfo {author} {\bibfnamefont {J.-P.}\ \bibnamefont
  {Blaizot}}, \bibinfo {author} {\bibfnamefont {H.}~\bibnamefont {Chate}},
  \bibinfo {author} {\bibfnamefont {B.}~\bibnamefont {Delamotte}}, \bibinfo
  {author} {\bibfnamefont {R.}~\bibnamefont {Mendez-Galain}}, \ and\ \bibinfo
  {author} {\bibfnamefont {N.}~\bibnamefont {Wschebor}},\ }\href {\doibase
  10.1103/PhysRevE.80.030103} {\bibfield  {journal} {\bibinfo  {journal} {Phys.
  Rev. E}\ }\textbf {\bibinfo {volume} {80}},\ \bibinfo {pages} {030103}
  (\bibinfo {year} {2009})},\ \Eprint {http://arxiv.org/abs/0901.0128}
  {arXiv:0901.0128 [cond-mat.stat-mech]} \BibitemShut {NoStop}%
\bibitem [{\citenamefont {Stokic}\ \emph {et~al.}(2010)\citenamefont {Stokic},
  \citenamefont {Friman},\ and\ \citenamefont {Redlich}}]{Stokic:2009uv}%
  \BibitemOpen
  \bibfield  {author} {\bibinfo {author} {\bibfnamefont {B.}~\bibnamefont
  {Stokic}}, \bibinfo {author} {\bibfnamefont {B.}~\bibnamefont {Friman}}, \
  and\ \bibinfo {author} {\bibfnamefont {K.}~\bibnamefont {Redlich}},\ }\href
  {\doibase 10.1140/epjc/s10052-010-1310-x} {\bibfield  {journal} {\bibinfo
  {journal} {Eur. Phys. J. C}\ }\textbf {\bibinfo {volume} {67}},\ \bibinfo
  {pages} {425} (\bibinfo {year} {2010})},\ \Eprint
  {http://arxiv.org/abs/0904.0466} {arXiv:0904.0466 [hep-ph]} \BibitemShut
  {NoStop}%
\bibitem [{\citenamefont {Litim}\ and\ \citenamefont
  {Zappala}(2011)}]{Litim:2010tt}%
  \BibitemOpen
  \bibfield  {author} {\bibinfo {author} {\bibfnamefont {D.~F.}\ \bibnamefont
  {Litim}}\ and\ \bibinfo {author} {\bibfnamefont {D.}~\bibnamefont
  {Zappala}},\ }\href {\doibase 10.1103/PhysRevD.83.085009} {\bibfield
  {journal} {\bibinfo  {journal} {Phys. Rev. D}\ }\textbf {\bibinfo {volume}
  {83}},\ \bibinfo {pages} {085009} (\bibinfo {year} {2011})},\ \Eprint
  {http://arxiv.org/abs/1009.1948} {arXiv:1009.1948 [hep-th]} \BibitemShut
  {NoStop}%
\bibitem [{\citenamefont {Benitez}\ \emph {et~al.}(2012)\citenamefont
  {Benitez}, \citenamefont {Blaizot}, \citenamefont {Chate}, \citenamefont
  {Delamotte}, \citenamefont {Mendez-Galain},\ and\ \citenamefont
  {Wschebor}}]{Benitez:2011xx}%
  \BibitemOpen
  \bibfield  {author} {\bibinfo {author} {\bibfnamefont {F.}~\bibnamefont
  {Benitez}}, \bibinfo {author} {\bibfnamefont {J.-P.}\ \bibnamefont
  {Blaizot}}, \bibinfo {author} {\bibfnamefont {H.}~\bibnamefont {Chate}},
  \bibinfo {author} {\bibfnamefont {B.}~\bibnamefont {Delamotte}}, \bibinfo
  {author} {\bibfnamefont {R.}~\bibnamefont {Mendez-Galain}}, \ and\ \bibinfo
  {author} {\bibfnamefont {N.}~\bibnamefont {Wschebor}},\ }\href {\doibase
  10.1103/PhysRevE.85.026707} {\bibfield  {journal} {\bibinfo  {journal} {Phys.
  Rev. E}\ }\textbf {\bibinfo {volume} {85}},\ \bibinfo {pages} {026707}
  (\bibinfo {year} {2012})},\ \Eprint {http://arxiv.org/abs/1110.2665}
  {arXiv:1110.2665 [cond-mat.stat-mech]} \BibitemShut {NoStop}%
\bibitem [{\citenamefont {Rancon}\ \emph {et~al.}(2013)\citenamefont {Rancon},
  \citenamefont {Kodio}, \citenamefont {Dupuis},\ and\ \citenamefont
  {Lecheminant}}]{Rancon_2013}%
  \BibitemOpen
  \bibfield  {author} {\bibinfo {author} {\bibfnamefont {A.}~\bibnamefont
  {Rancon}}, \bibinfo {author} {\bibfnamefont {O.}~\bibnamefont {Kodio}},
  \bibinfo {author} {\bibfnamefont {N.}~\bibnamefont {Dupuis}}, \ and\ \bibinfo
  {author} {\bibfnamefont {P.}~\bibnamefont {Lecheminant}},\ }\href {\doibase
  10.1103/physreve.88.012113} {\bibfield  {journal} {\bibinfo  {journal}
  {Physical Review E}\ }\textbf {\bibinfo {volume} {88}} (\bibinfo {year}
  {2013}),\ 10.1103/physreve.88.012113}\BibitemShut {NoStop}%
\bibitem [{\citenamefont {Defenu}\ \emph {et~al.}(2015)\citenamefont {Defenu},
  \citenamefont {Trombettoni},\ and\ \citenamefont {Codello}}]{Defenu:2014bea}%
  \BibitemOpen
  \bibfield  {author} {\bibinfo {author} {\bibfnamefont {N.}~\bibnamefont
  {Defenu}}, \bibinfo {author} {\bibfnamefont {A.}~\bibnamefont {Trombettoni}},
  \ and\ \bibinfo {author} {\bibfnamefont {A.}~\bibnamefont {Codello}},\ }\href
  {\doibase 10.1103/PhysRevE.92.052113} {\bibfield  {journal} {\bibinfo
  {journal} {Phys. Rev. E}\ }\textbf {\bibinfo {volume} {92}},\ \bibinfo
  {pages} {052113} (\bibinfo {year} {2015})},\ \Eprint
  {http://arxiv.org/abs/1409.8322} {arXiv:1409.8322 [cond-mat.stat-mech]}
  \BibitemShut {NoStop}%
\bibitem [{\citenamefont {Codello}\ \emph {et~al.}(2015)\citenamefont
  {Codello}, \citenamefont {Defenu},\ and\ \citenamefont
  {D'Odorico}}]{Codello:2014yfa}%
  \BibitemOpen
  \bibfield  {author} {\bibinfo {author} {\bibfnamefont {A.}~\bibnamefont
  {Codello}}, \bibinfo {author} {\bibfnamefont {N.}~\bibnamefont {Defenu}}, \
  and\ \bibinfo {author} {\bibfnamefont {G.}~\bibnamefont {D'Odorico}},\ }\href
  {\doibase 10.1103/PhysRevD.91.105003} {\bibfield  {journal} {\bibinfo
  {journal} {Phys. Rev. D}\ }\textbf {\bibinfo {volume} {91}},\ \bibinfo
  {pages} {105003} (\bibinfo {year} {2015})},\ \Eprint
  {http://arxiv.org/abs/1410.3308} {arXiv:1410.3308 [hep-th]} \BibitemShut
  {NoStop}%
\bibitem [{\citenamefont {Eichhorn}\ \emph {et~al.}(2016)\citenamefont
  {Eichhorn}, \citenamefont {Janssen},\ and\ \citenamefont
  {Scherer}}]{Eichhorn:2016hdi}%
  \BibitemOpen
  \bibfield  {author} {\bibinfo {author} {\bibfnamefont {A.}~\bibnamefont
  {Eichhorn}}, \bibinfo {author} {\bibfnamefont {L.}~\bibnamefont {Janssen}}, \
  and\ \bibinfo {author} {\bibfnamefont {M.~M.}\ \bibnamefont {Scherer}},\
  }\href {\doibase 10.1103/PhysRevD.93.125021} {\bibfield  {journal} {\bibinfo
  {journal} {Phys. Rev. D}\ }\textbf {\bibinfo {volume} {93}},\ \bibinfo
  {pages} {125021} (\bibinfo {year} {2016})},\ \Eprint
  {http://arxiv.org/abs/1604.03561} {arXiv:1604.03561 [hep-th]} \BibitemShut
  {NoStop}%
\bibitem [{\citenamefont {Litim}\ and\ \citenamefont
  {Marchais}(2017)}]{Litim:2016hlb}%
  \BibitemOpen
  \bibfield  {author} {\bibinfo {author} {\bibfnamefont {D.~F.}\ \bibnamefont
  {Litim}}\ and\ \bibinfo {author} {\bibfnamefont {E.}~\bibnamefont
  {Marchais}},\ }\href {\doibase 10.1103/PhysRevD.95.025026} {\bibfield
  {journal} {\bibinfo  {journal} {Phys. Rev. D}\ }\textbf {\bibinfo {volume}
  {95}},\ \bibinfo {pages} {025026} (\bibinfo {year} {2017})},\ \Eprint
  {http://arxiv.org/abs/1607.02030} {arXiv:1607.02030 [hep-th]} \BibitemShut
  {NoStop}%
\bibitem [{\citenamefont {Jüttner}\ \emph {et~al.}(2017)\citenamefont
  {Jüttner}, \citenamefont {Litim},\ and\ \citenamefont
  {Marchais}}]{Juttner:2017cpr}%
  \BibitemOpen
  \bibfield  {author} {\bibinfo {author} {\bibfnamefont {A.}~\bibnamefont
  {Jüttner}}, \bibinfo {author} {\bibfnamefont {D.~F.}\ \bibnamefont {Litim}},
  \ and\ \bibinfo {author} {\bibfnamefont {E.}~\bibnamefont {Marchais}},\
  }\href {\doibase 10.1016/j.nuclphysb.2017.06.010} {\bibfield  {journal}
  {\bibinfo  {journal} {Nucl. Phys. B}\ }\textbf {\bibinfo {volume} {921}},\
  \bibinfo {pages} {769} (\bibinfo {year} {2017})},\ \Eprint
  {http://arxiv.org/abs/1701.05168} {arXiv:1701.05168 [hep-th]} \BibitemShut
  {NoStop}%
\bibitem [{\citenamefont {Roscher}\ and\ \citenamefont
  {Herbut}(2018)}]{Roscher:2018ucp}%
  \BibitemOpen
  \bibfield  {author} {\bibinfo {author} {\bibfnamefont {D.}~\bibnamefont
  {Roscher}}\ and\ \bibinfo {author} {\bibfnamefont {I.~F.}\ \bibnamefont
  {Herbut}},\ }\href {\doibase 10.1103/PhysRevD.97.116019} {\bibfield
  {journal} {\bibinfo  {journal} {Phys. Rev. D}\ }\textbf {\bibinfo {volume}
  {97}},\ \bibinfo {pages} {116019} (\bibinfo {year} {2018})},\ \Eprint
  {http://arxiv.org/abs/1805.01480} {arXiv:1805.01480 [hep-th]} \BibitemShut
  {NoStop}%
\bibitem [{\citenamefont {Yabunaka}\ and\ \citenamefont
  {Delamotte}(2018)}]{Yabunaka:2018mju}%
  \BibitemOpen
  \bibfield  {author} {\bibinfo {author} {\bibfnamefont {S.}~\bibnamefont
  {Yabunaka}}\ and\ \bibinfo {author} {\bibfnamefont {B.}~\bibnamefont
  {Delamotte}},\ }\href {\doibase 10.1103/PhysRevLett.121.231601} {\bibfield
  {journal} {\bibinfo  {journal} {Phys. Rev. Lett.}\ }\textbf {\bibinfo
  {volume} {121}},\ \bibinfo {pages} {231601} (\bibinfo {year} {2018})},\
  \Eprint {http://arxiv.org/abs/1807.04681} {arXiv:1807.04681
  [cond-mat.stat-mech]} \BibitemShut {NoStop}%
\bibitem [{\citenamefont {De~Polsi}\ \emph {et~al.}(2020)\citenamefont
  {De~Polsi}, \citenamefont {Balog}, \citenamefont {Tissier},\ and\
  \citenamefont {Wschebor}}]{DePolsi:2020pjk}%
  \BibitemOpen
  \bibfield  {author} {\bibinfo {author} {\bibfnamefont {G.}~\bibnamefont
  {De~Polsi}}, \bibinfo {author} {\bibfnamefont {I.}~\bibnamefont {Balog}},
  \bibinfo {author} {\bibfnamefont {M.}~\bibnamefont {Tissier}}, \ and\
  \bibinfo {author} {\bibfnamefont {N.}~\bibnamefont {Wschebor}},\ }\href
  {\doibase 10.1103/PhysRevE.101.042113} {\bibfield  {journal} {\bibinfo
  {journal} {Phys. Rev. E}\ }\textbf {\bibinfo {volume} {101}},\ \bibinfo
  {pages} {042113} (\bibinfo {year} {2020})},\ \Eprint
  {http://arxiv.org/abs/2001.07525} {arXiv:2001.07525 [cond-mat.stat-mech]}
  \BibitemShut {NoStop}%
\bibitem [{\citenamefont {Morris}(1994)}]{Morris:1994ie}%
  \BibitemOpen
  \bibfield  {author} {\bibinfo {author} {\bibfnamefont {T.~R.}\ \bibnamefont
  {Morris}},\ }\href {\doibase 10.1016/0370-2693(94)90767-6} {\bibfield
  {journal} {\bibinfo  {journal} {Phys. Lett. B}\ }\textbf {\bibinfo {volume}
  {329}},\ \bibinfo {pages} {241} (\bibinfo {year} {1994})},\ \Eprint
  {http://arxiv.org/abs/hep-ph/9403340} {arXiv:hep-ph/9403340} \BibitemShut
  {NoStop}%
\bibitem [{\citenamefont {Balog}\ \emph {et~al.}(2019)\citenamefont {Balog},
  \citenamefont {Chaté}, \citenamefont {Delamotte}, \citenamefont {Marohnic},\
  and\ \citenamefont {Wschebor}}]{Balog:2019rrg}%
  \BibitemOpen
  \bibfield  {author} {\bibinfo {author} {\bibfnamefont {I.}~\bibnamefont
  {Balog}}, \bibinfo {author} {\bibfnamefont {H.}~\bibnamefont {Chaté}},
  \bibinfo {author} {\bibfnamefont {B.}~\bibnamefont {Delamotte}}, \bibinfo
  {author} {\bibfnamefont {M.}~\bibnamefont {Marohnic}}, \ and\ \bibinfo
  {author} {\bibfnamefont {N.}~\bibnamefont {Wschebor}},\ }\href {\doibase
  10.1103/PhysRevLett.123.240604} {\bibfield  {journal} {\bibinfo  {journal}
  {Phys. Rev. Lett.}\ }\textbf {\bibinfo {volume} {123}},\ \bibinfo {pages}
  {240604} (\bibinfo {year} {2019})},\ \Eprint
  {http://arxiv.org/abs/1907.01829} {arXiv:1907.01829 [cond-mat.stat-mech]}
  \BibitemShut {NoStop}%
\bibitem [{\citenamefont {Litim}(2000)}]{Litim:2000ci}%
  \BibitemOpen
  \bibfield  {author} {\bibinfo {author} {\bibfnamefont {D.~F.}\ \bibnamefont
  {Litim}},\ }\href {\doibase 10.1016/S0370-2693(00)00748-6} {\bibfield
  {journal} {\bibinfo  {journal} {Phys. Lett. B}\ }\textbf {\bibinfo {volume}
  {486}},\ \bibinfo {pages} {92} (\bibinfo {year} {2000})},\ \Eprint
  {http://arxiv.org/abs/hep-th/0005245} {arXiv:hep-th/0005245} \BibitemShut
  {NoStop}%
\bibitem [{\citenamefont {Litim}(2001{\natexlab{b}})}]{Litim:2001up}%
  \BibitemOpen
  \bibfield  {author} {\bibinfo {author} {\bibfnamefont {D.~F.}\ \bibnamefont
  {Litim}},\ }\href {\doibase 10.1103/PhysRevD.64.105007} {\bibfield  {journal}
  {\bibinfo  {journal} {Phys. Rev.}\ }\textbf {\bibinfo {volume} {D64}},\
  \bibinfo {pages} {105007} (\bibinfo {year} {2001}{\natexlab{b}})},\ \Eprint
  {http://arxiv.org/abs/hep-th/0103195} {arXiv:hep-th/0103195 [hep-th]}
  \BibitemShut {NoStop}%
\bibitem [{\citenamefont {Litim}(2001{\natexlab{c}})}]{Litim:2001fd}%
  \BibitemOpen
  \bibfield  {author} {\bibinfo {author} {\bibfnamefont {D.~F.}\ \bibnamefont
  {Litim}},\ }\href {\doibase 10.1142/S0217751X01004748} {\bibfield  {journal}
  {\bibinfo  {journal} {Int. J. Mod. Phys. A}\ }\textbf {\bibinfo {volume}
  {16}},\ \bibinfo {pages} {2081} (\bibinfo {year} {2001}{\natexlab{c}})},\
  \Eprint {http://arxiv.org/abs/hep-th/0104221} {arXiv:hep-th/0104221}
  \BibitemShut {NoStop}%
\bibitem [{\citenamefont {Canet}\ \emph {et~al.}(2003)\citenamefont {Canet},
  \citenamefont {Delamotte}, \citenamefont {Mouhanna},\ and\ \citenamefont
  {Vidal}}]{Canet:2002gs}%
  \BibitemOpen
  \bibfield  {author} {\bibinfo {author} {\bibfnamefont {L.}~\bibnamefont
  {Canet}}, \bibinfo {author} {\bibfnamefont {B.}~\bibnamefont {Delamotte}},
  \bibinfo {author} {\bibfnamefont {D.}~\bibnamefont {Mouhanna}}, \ and\
  \bibinfo {author} {\bibfnamefont {J.}~\bibnamefont {Vidal}},\ }\href
  {\doibase 10.1103/PhysRevD.67.065004} {\bibfield  {journal} {\bibinfo
  {journal} {Phys. Rev. D}\ }\textbf {\bibinfo {volume} {67}},\ \bibinfo
  {pages} {065004} (\bibinfo {year} {2003})},\ \Eprint
  {http://arxiv.org/abs/hep-th/0211055} {arXiv:hep-th/0211055} \BibitemShut
  {NoStop}%
\bibitem [{\citenamefont {Pawlowski}(2007)}]{Pawlowski:2005xe}%
  \BibitemOpen
  \bibfield  {author} {\bibinfo {author} {\bibfnamefont {J.~M.}\ \bibnamefont
  {Pawlowski}},\ }\href {\doibase 10.1016/j.aop.2007.01.007} {\bibfield
  {journal} {\bibinfo  {journal} {Annals Phys.}\ }\textbf {\bibinfo {volume}
  {322}},\ \bibinfo {pages} {2831} (\bibinfo {year} {2007})},\ \Eprint
  {http://arxiv.org/abs/hep-th/0512261} {arXiv:hep-th/0512261} \BibitemShut
  {NoStop}%
\bibitem [{\citenamefont {Litim}(2007)}]{Litim:2007jb}%
  \BibitemOpen
  \bibfield  {author} {\bibinfo {author} {\bibfnamefont {D.~F.}\ \bibnamefont
  {Litim}},\ }\href {\doibase 10.1103/PhysRevD.76.105001} {\bibfield  {journal}
  {\bibinfo  {journal} {Phys. Rev. D}\ }\textbf {\bibinfo {volume} {76}},\
  \bibinfo {pages} {105001} (\bibinfo {year} {2007})},\ \Eprint
  {http://arxiv.org/abs/0704.1514} {arXiv:0704.1514 [hep-th]} \BibitemShut
  {NoStop}%
\bibitem [{\citenamefont {Pawlowski}\ \emph {et~al.}(2017)\citenamefont
  {Pawlowski}, \citenamefont {Scherer}, \citenamefont {Schmidt},\ and\
  \citenamefont {Wetzel}}]{Pawlowski:2015mlf}%
  \BibitemOpen
  \bibfield  {author} {\bibinfo {author} {\bibfnamefont {J.~M.}\ \bibnamefont
  {Pawlowski}}, \bibinfo {author} {\bibfnamefont {M.~M.}\ \bibnamefont
  {Scherer}}, \bibinfo {author} {\bibfnamefont {R.}~\bibnamefont {Schmidt}}, \
  and\ \bibinfo {author} {\bibfnamefont {S.~J.}\ \bibnamefont {Wetzel}},\
  }\href {\doibase 10.1016/j.aop.2017.06.017} {\bibfield  {journal} {\bibinfo
  {journal} {Annals Phys.}\ }\textbf {\bibinfo {volume} {384}},\ \bibinfo
  {pages} {165} (\bibinfo {year} {2017})},\ \Eprint
  {http://arxiv.org/abs/1512.03598} {arXiv:1512.03598 [hep-th]} \BibitemShut
  {NoStop}%
\bibitem [{\citenamefont {Tetradis}\ and\ \citenamefont
  {Wetterich}(1994)}]{Tetradis:1993bx}%
  \BibitemOpen
  \bibfield  {author} {\bibinfo {author} {\bibfnamefont {N.}~\bibnamefont
  {Tetradis}}\ and\ \bibinfo {author} {\bibfnamefont {C.}~\bibnamefont
  {Wetterich}},\ }\href {\doibase 10.1142/S0217751X94001631} {\bibfield
  {journal} {\bibinfo  {journal} {Int. J. Mod. Phys. A}\ }\textbf {\bibinfo
  {volume} {9}},\ \bibinfo {pages} {4029} (\bibinfo {year} {1994})},\ \Eprint
  {http://arxiv.org/abs/hep-ph/9309257} {arXiv:hep-ph/9309257} \BibitemShut
  {NoStop}%
\bibitem [{\citenamefont {Litim}\ and\ \citenamefont
  {Pawlowski}(2006)}]{Litim:2006ag}%
  \BibitemOpen
  \bibfield  {author} {\bibinfo {author} {\bibfnamefont {D.~F.}\ \bibnamefont
  {Litim}}\ and\ \bibinfo {author} {\bibfnamefont {J.~M.}\ \bibnamefont
  {Pawlowski}},\ }\href {\doibase 10.1088/1126-6708/2006/11/026} {\bibfield
  {journal} {\bibinfo  {journal} {JHEP}\ }\textbf {\bibinfo {volume} {11}},\
  \bibinfo {pages} {026} (\bibinfo {year} {2006})},\ \Eprint
  {http://arxiv.org/abs/hep-th/0609122} {arXiv:hep-th/0609122} \BibitemShut
  {NoStop}%
\bibitem [{\citenamefont {Pawlowski}\ and\ \citenamefont
  {Rennecke}(2014)}]{Paw}%
  \BibitemOpen
  \bibfield  {author} {\bibinfo {author} {\bibfnamefont {J.~M.}\ \bibnamefont
  {Pawlowski}}\ and\ \bibinfo {author} {\bibfnamefont {F.}~\bibnamefont
  {Rennecke}},\ }\href {\doibase 10.1103/PhysRevD.90.076002} {\bibfield
  {journal} {\bibinfo  {journal} {Phys. Rev.}\ }\textbf {\bibinfo {volume}
  {D90}},\ \bibinfo {pages} {076002} (\bibinfo {year} {2014})},\ \Eprint
  {http://arxiv.org/abs/1403.1179} {arXiv:1403.1179 [hep-ph]} \BibitemShut
  {NoStop}%
\bibitem [{Note2()}]{Note2}%
  \BibitemOpen
  \bibinfo {note} {We checked that our numerical calculations reproduce the
  O(4) magnetic equation of state of Ref.~\cite {Braun:2007td, Engels:2011km}
  for real values of the argument.}\BibitemShut {Stop}%
\bibitem [{\citenamefont {Kos}\ \emph {et~al.}(2016)\citenamefont {Kos},
  \citenamefont {Poland}, \citenamefont {Simmons-Duffin},\ and\ \citenamefont
  {Vichi}}]{Kos:2016ysd}%
  \BibitemOpen
  \bibfield  {author} {\bibinfo {author} {\bibfnamefont {F.}~\bibnamefont
  {Kos}}, \bibinfo {author} {\bibfnamefont {D.}~\bibnamefont {Poland}},
  \bibinfo {author} {\bibfnamefont {D.}~\bibnamefont {Simmons-Duffin}}, \ and\
  \bibinfo {author} {\bibfnamefont {A.}~\bibnamefont {Vichi}},\ }\href
  {\doibase 10.1007/JHEP08(2016)036} {\bibfield  {journal} {\bibinfo  {journal}
  {JHEP}\ }\textbf {\bibinfo {volume} {08}},\ \bibinfo {pages} {036} (\bibinfo
  {year} {2016})},\ \Eprint {http://arxiv.org/abs/1603.04436} {arXiv:1603.04436
  [hep-th]} \BibitemShut {NoStop}%
\bibitem [{\citenamefont {Johnson}\ \emph {et~al.}()\citenamefont {Johnson},
  \citenamefont {Rennecke},\ and\ \citenamefont {Skokov}}]{CRS2020}%
  \BibitemOpen
  \bibfield  {author} {\bibinfo {author} {\bibfnamefont {G.}~\bibnamefont
  {Johnson}}, \bibinfo {author} {\bibfnamefont {F.}~\bibnamefont {Rennecke}}, \
  and\ \bibinfo {author} {\bibfnamefont {V.}~\bibnamefont {Skokov}},\
  }\href@noop {} {}\bibinfo {note} {To be submitted}\BibitemShut {NoStop}%
\bibitem [{\citenamefont {Brezin}\ and\ \citenamefont
  {Wallace}(1973)}]{Brezin:1972se}%
  \BibitemOpen
  \bibfield  {author} {\bibinfo {author} {\bibfnamefont {E.}~\bibnamefont
  {Brezin}}\ and\ \bibinfo {author} {\bibfnamefont {D.}~\bibnamefont
  {Wallace}},\ }\href {\doibase 10.1103/PhysRevB.7.1967} {\bibfield  {journal}
  {\bibinfo  {journal} {Phys. Rev. B}\ }\textbf {\bibinfo {volume} {7}},\
  \bibinfo {pages} {1967} (\bibinfo {year} {1973})}\BibitemShut {NoStop}%
\bibitem [{\citenamefont {Moshe}\ and\ \citenamefont
  {Zinn-Justin}(2003)}]{Moshe:2003xn}%
  \BibitemOpen
  \bibfield  {author} {\bibinfo {author} {\bibfnamefont {M.}~\bibnamefont
  {Moshe}}\ and\ \bibinfo {author} {\bibfnamefont {J.}~\bibnamefont
  {Zinn-Justin}},\ }\href {\doibase 10.1016/S0370-1573(03)00263-1} {\bibfield
  {journal} {\bibinfo  {journal} {Phys. Rept.}\ }\textbf {\bibinfo {volume}
  {385}},\ \bibinfo {pages} {69} (\bibinfo {year} {2003})},\ \Eprint
  {http://arxiv.org/abs/hep-th/0306133} {arXiv:hep-th/0306133} \BibitemShut
  {NoStop}%
\bibitem [{\citenamefont {Almasi}\ \emph {et~al.}(2017)\citenamefont {Almasi},
  \citenamefont {Tarnowski}, \citenamefont {Friman},\ and\ \citenamefont
  {Redlich}}]{Almasi:2016gcn}%
  \BibitemOpen
  \bibfield  {author} {\bibinfo {author} {\bibfnamefont {G.~A.}\ \bibnamefont
  {Almasi}}, \bibinfo {author} {\bibfnamefont {W.}~\bibnamefont {Tarnowski}},
  \bibinfo {author} {\bibfnamefont {B.}~\bibnamefont {Friman}}, \ and\ \bibinfo
  {author} {\bibfnamefont {K.}~\bibnamefont {Redlich}},\ }\href {\doibase
  10.1103/PhysRevD.95.014007} {\bibfield  {journal} {\bibinfo  {journal} {Phys.
  Rev. D}\ }\textbf {\bibinfo {volume} {95}},\ \bibinfo {pages} {014007}
  (\bibinfo {year} {2017})},\ \Eprint {http://arxiv.org/abs/1605.05423}
  {arXiv:1605.05423 [hep-ph]} \BibitemShut {NoStop}%
\bibitem [{\citenamefont {Henrici}(1991)}]{henrici1991applied}%
  \BibitemOpen
  \bibfield  {author} {\bibinfo {author} {\bibfnamefont {P.}~\bibnamefont
  {Henrici}},\ }\href {https://books.google.com/books?id=KPpQAAAAMAAJ} {\emph
  {\bibinfo {title} {Applied and Computational Complex Analysis, Volume 2:
  Special Functions, Integral Transforms, Asymptotics, Continued Fractions}}},\
  Applied and Computational Complex Analysis\ (\bibinfo  {publisher} {Wiley},\
  \bibinfo {year} {1991})\BibitemShut {NoStop}%
\bibitem [{\citenamefont {Bazavov}\ \emph {et~al.}(2019)\citenamefont {Bazavov}
  \emph {et~al.}}]{Bazavov:2018mes}%
  \BibitemOpen
  \bibfield  {author} {\bibinfo {author} {\bibfnamefont {A.}~\bibnamefont
  {Bazavov}} \emph {et~al.} (\bibinfo {collaboration} {HotQCD}),\ }\href
  {\doibase 10.1016/j.physletb.2019.05.013} {\bibfield  {journal} {\bibinfo
  {journal} {Phys. Lett.}\ }\textbf {\bibinfo {volume} {B795}},\ \bibinfo
  {pages} {15} (\bibinfo {year} {2019})},\ \Eprint
  {http://arxiv.org/abs/1812.08235} {arXiv:1812.08235 [hep-lat]} \BibitemShut
  {NoStop}%
\bibitem [{\citenamefont {Borsanyi}\ \emph {et~al.}(2019)\citenamefont
  {Borsanyi}, \citenamefont {Fodor}, \citenamefont {Guenther}, \citenamefont
  {Katz}, \citenamefont {Pasztor}, \citenamefont {Portillo}, \citenamefont
  {Ratti},\ and\ \citenamefont {Szabó}}]{Borsanyi:2019hsj}%
  \BibitemOpen
  \bibfield  {author} {\bibinfo {author} {\bibfnamefont {S.}~\bibnamefont
  {Borsanyi}}, \bibinfo {author} {\bibfnamefont {Z.}~\bibnamefont {Fodor}},
  \bibinfo {author} {\bibfnamefont {J.~N.}\ \bibnamefont {Guenther}}, \bibinfo
  {author} {\bibfnamefont {S.~K.}\ \bibnamefont {Katz}}, \bibinfo {author}
  {\bibfnamefont {A.}~\bibnamefont {Pasztor}}, \bibinfo {author} {\bibfnamefont
  {I.}~\bibnamefont {Portillo}}, \bibinfo {author} {\bibfnamefont
  {C.}~\bibnamefont {Ratti}}, \ and\ \bibinfo {author} {\bibfnamefont {K.~K.}\
  \bibnamefont {Szabó}},\ }\href {\doibase 10.1016/j.nuclphysa.2018.12.016}
  {\bibfield  {journal} {\bibinfo  {journal} {Nucl. Phys.}\ }\textbf {\bibinfo
  {volume} {A982}},\ \bibinfo {pages} {223} (\bibinfo {year}
  {2019})}\BibitemShut {NoStop}%
\bibitem [{\citenamefont {Fischer}(2019)}]{Fischer:2018sdj}%
  \BibitemOpen
  \bibfield  {author} {\bibinfo {author} {\bibfnamefont {C.~S.}\ \bibnamefont
  {Fischer}},\ }\href {\doibase 10.1016/j.ppnp.2019.01.002} {\bibfield
  {journal} {\bibinfo  {journal} {Prog. Part. Nucl. Phys.}\ }\textbf {\bibinfo
  {volume} {105}},\ \bibinfo {pages} {1} (\bibinfo {year} {2019})},\ \Eprint
  {http://arxiv.org/abs/1810.12938} {arXiv:1810.12938 [hep-ph]} \BibitemShut
  {NoStop}%
\bibitem [{\citenamefont {Fu}\ \emph {et~al.}(2020)\citenamefont {Fu},
  \citenamefont {Pawlowski},\ and\ \citenamefont {Rennecke}}]{Fu:2019hdw}%
  \BibitemOpen
  \bibfield  {author} {\bibinfo {author} {\bibfnamefont {W.-j.}\ \bibnamefont
  {Fu}}, \bibinfo {author} {\bibfnamefont {J.~M.}\ \bibnamefont {Pawlowski}}, \
  and\ \bibinfo {author} {\bibfnamefont {F.}~\bibnamefont {Rennecke}},\ }\href
  {\doibase 10.1103/PhysRevD.101.054032} {\bibfield  {journal} {\bibinfo
  {journal} {Phys. Rev. D}\ }\textbf {\bibinfo {volume} {101}},\ \bibinfo
  {pages} {054032} (\bibinfo {year} {2020})},\ \Eprint
  {http://arxiv.org/abs/1909.02991} {arXiv:1909.02991 [hep-ph]} \BibitemShut
  {NoStop}%
\bibitem [{\citenamefont {Gao}\ and\ \citenamefont
  {Pawlowski}(2020)}]{Gao:2020qsj}%
  \BibitemOpen
  \bibfield  {author} {\bibinfo {author} {\bibfnamefont {F.}~\bibnamefont
  {Gao}}\ and\ \bibinfo {author} {\bibfnamefont {J.~M.}\ \bibnamefont
  {Pawlowski}},\ }\href {\doibase 10.1103/PhysRevD.102.034027} {\bibfield
  {journal} {\bibinfo  {journal} {Phys. Rev. D}\ }\textbf {\bibinfo {volume}
  {102}},\ \bibinfo {pages} {034027} (\bibinfo {year} {2020})},\ \Eprint
  {http://arxiv.org/abs/2002.07500} {arXiv:2002.07500 [hep-ph]} \BibitemShut
  {NoStop}%
\bibitem [{\citenamefont {Nonaka}\ and\ \citenamefont
  {Asakawa}(2005)}]{Nonaka:2004pg}%
  \BibitemOpen
  \bibfield  {author} {\bibinfo {author} {\bibfnamefont {C.}~\bibnamefont
  {Nonaka}}\ and\ \bibinfo {author} {\bibfnamefont {M.}~\bibnamefont
  {Asakawa}},\ }\href {\doibase 10.1103/PhysRevC.71.044904} {\bibfield
  {journal} {\bibinfo  {journal} {Phys. Rev. C}\ }\textbf {\bibinfo {volume}
  {71}},\ \bibinfo {pages} {044904} (\bibinfo {year} {2005})},\ \Eprint
  {http://arxiv.org/abs/nucl-th/0410078} {arXiv:nucl-th/0410078} \BibitemShut
  {NoStop}%
\bibitem [{\citenamefont {Stephanov}(2011)}]{Stephanov:2011pb}%
  \BibitemOpen
  \bibfield  {author} {\bibinfo {author} {\bibfnamefont {M.}~\bibnamefont
  {Stephanov}},\ }\href {\doibase 10.1103/PhysRevLett.107.052301} {\bibfield
  {journal} {\bibinfo  {journal} {Phys. Rev. Lett.}\ }\textbf {\bibinfo
  {volume} {107}},\ \bibinfo {pages} {052301} (\bibinfo {year} {2011})},\
  \Eprint {http://arxiv.org/abs/1104.1627} {arXiv:1104.1627 [hep-ph]}
  \BibitemShut {NoStop}%
\bibitem [{\citenamefont {Parotto}\ \emph {et~al.}(2020)\citenamefont
  {Parotto}, \citenamefont {Bluhm}, \citenamefont {Mroczek}, \citenamefont
  {Nahrgang}, \citenamefont {Noronha-Hostler}, \citenamefont {Rajagopal},
  \citenamefont {Ratti}, \citenamefont {Sch\"afer},\ and\ \citenamefont
  {Stephanov}}]{Parotto:2018pwx}%
  \BibitemOpen
  \bibfield  {author} {\bibinfo {author} {\bibfnamefont {P.}~\bibnamefont
  {Parotto}}, \bibinfo {author} {\bibfnamefont {M.}~\bibnamefont {Bluhm}},
  \bibinfo {author} {\bibfnamefont {D.}~\bibnamefont {Mroczek}}, \bibinfo
  {author} {\bibfnamefont {M.}~\bibnamefont {Nahrgang}}, \bibinfo {author}
  {\bibfnamefont {J.}~\bibnamefont {Noronha-Hostler}}, \bibinfo {author}
  {\bibfnamefont {K.}~\bibnamefont {Rajagopal}}, \bibinfo {author}
  {\bibfnamefont {C.}~\bibnamefont {Ratti}}, \bibinfo {author} {\bibfnamefont
  {T.}~\bibnamefont {Sch\"afer}}, \ and\ \bibinfo {author} {\bibfnamefont
  {M.}~\bibnamefont {Stephanov}},\ }\href {\doibase
  10.1103/PhysRevC.101.034901} {\bibfield  {journal} {\bibinfo  {journal}
  {Phys. Rev. C}\ }\textbf {\bibinfo {volume} {101}},\ \bibinfo {pages}
  {034901} (\bibinfo {year} {2020})},\ \Eprint
  {http://arxiv.org/abs/1805.05249} {arXiv:1805.05249 [hep-ph]} \BibitemShut
  {NoStop}%
\bibitem [{\citenamefont {Engels}\ and\ \citenamefont
  {Karsch}(2012)}]{Engels:2011km}%
  \BibitemOpen
  \bibfield  {author} {\bibinfo {author} {\bibfnamefont {J.}~\bibnamefont
  {Engels}}\ and\ \bibinfo {author} {\bibfnamefont {F.}~\bibnamefont
  {Karsch}},\ }\href {\doibase 10.1103/PhysRevD.85.094506} {\bibfield
  {journal} {\bibinfo  {journal} {Phys. Rev. D}\ }\textbf {\bibinfo {volume}
  {85}},\ \bibinfo {pages} {094506} (\bibinfo {year} {2012})},\ \Eprint
  {http://arxiv.org/abs/1105.0584} {arXiv:1105.0584 [hep-lat]} \BibitemShut
  {NoStop}%
\end{thebibliography}%

\end{document}